\input{aipcheck}

\documentclass[
    ,final            
  ]
  {aipproc}

\layoutstyle{8x11single}

\def\gtap{\ \raise.3ex\hbox{$>$\kern-.75em\lower1ex\hbox{$\sim$}}\ }
\def\ltap{\ \raise.3ex\hbox{$<$\kern-.75em\lower1ex\hbox{$\sim$}}\ }
\usepackage{bm}

\newcommand{\eqn}[1]{\label{eq:#1}}
\newcommand{\refeq}[1]{(\ref{eq:#1})}

\newcommand{\Eq}{Eq.~\refeq}

\begin{document}

\title{Dynamical Model of Coherent Electroweak Pion Production
Reactions on Nuclei }


\author{S. X. Nakamura}{
  address={Instituto de F\'isica, Universidade de S\~ao Paulo,
C.P. 66318, 05315-970, S\~ao Paulo, SP, Brazil}
}

\author{T. Sato}{
  address={Department of Physics, Osaka University, Toyonaka, Osaka,
 560-0043, Japan}
}

\author{T.-S. H. Lee}{
  address={Physics Division, Argonne National Laboratory, Argonne, Illinois 60439, USA}
}

\author{B. Szczerbinska}{
  address={Dakota State University, College of Arts \& Sciences,
Madison, SD, 57042-1799, USA}
}

\author{K. Kubodera}{
  address={Department of Physics and Astronomy, University of South
 Carolina, Columbia, SC, 29208, USA}
}

\begin{abstract}
We have developed a dynamical model for a unified description of the
 pion-nucleus scattering and photo- and neutrino-induced coherent pion
production on nuclei.
Our approach is based on a combined use of the Sato-Lee model 
for the  electroweak pion production on a single nucleon
and the $\Delta$-hole model of pion-nucleus scattering.
Numerical calculations are carried out for the case of the $^{12}$C
target.
After testing our model with the use of  the pion photo-production data,
we confront our predictions of the neutrino-induced coherent pion production
reactions with
the recent data from K2K and MiniBooNE.
\end{abstract}

\maketitle


\section{Introduction}

In recent years both the theoretical and experimental studies on the
coherent pion production in neutrino-nucleus reactions
in few-GeV region have been intense.
These studies were  originally inspired by the on-going and future
neutrino experiments.
Some interesting results from recent experiments have 
further encouraged the theoretical studies of this process.
Namely, the K2K\cite{hasegawa} and SciBooNE\cite{hiraide} collaborations reported no evidence of the
charged-current (CC) process, while the MiniBooNE collaboration found a 
signal of the neutral-current (NC) process.\cite{miniboone}
This discrepancy must be resolved since  the isospin consideration
predicts  the simple relation
$\sigma_{CC}\sim 2\sigma_{NC}$.
The on-going data analysis of the coherent process have been trying to
provide more detailed information including the muon and pion
kinematics for the CC process,  and some preliminary results have been
presented in this workshop\cite{hiraide_nuint09,tanaka_nuint09}.
%
%
Considering these experimental developments, it is important
to develop a rigorous theoretical model for analyzing the available
and the forthcoming data.
This is the objective of our work.

There are mainly two types of theoretical approaches to the coherent pion
production, i.e., a PCAC-based model and a microscopic model. We pursue
the latter option.
In a microscopic  approach, the starting point is 
a model which can  describe well the
elementary electroweak pion production on a single nucleon
 in the considered energy region.
Such a model has been developed by Sato and Lee\cite{sl} (SL) and
is used in this work.
The SL model also provides accurate pion-nucleon scattering amplitudes
which are derived in a way consistent with
the pion production amplitudes from the same Lagrangian. 
The $\pi N$ scattering amplitudes
are  important ingredients in our
calculations of an optical potential which describes the
essential final pion-nucleus interactions.  
This optical potential takes account of the most important
medium modifications of the 
$\Delta$-propagation within the well-established 
 $\Delta$-hole model of pion-nucleus scattering.
Therefore, a combined use of the SL model and the $\Delta$-hole model,
which seems promising for calculating pion production in nuclei, 
provide us with a consistent microscopic description of the
pion scattering and the electroweak pion production mechanisms in nuclei. 
This consistency is an appealing point of our approach over the previous
microscopic models for the coherent pion productions, and
enables us to: (i) fix all free parameters in our model
using the pion-nucleus scattering data; (ii) perform parameter-free
calculations for the coherent pion productions; (iii) test the
reliability of our approach using data for the photo-production process.
Another important point to be noted is 
that we take care of the non-local effects
on the in-medium $\Delta$-propagation
whose possibly large effect was pointed out recently~\cite{non-local}.
For the neutrino-induced coherent pion production, 
no previous calculations (either in the PCAC-based or microscopic model) 
have included this effect.

Our calculations proceed as follows.
We first employ the on- and off-shell $\pi N$ scattering amplitudes generated
from the SL model to construct a pion-nucleus optical potential within
the $\Delta$-hole model\cite{karaoglu}.
The parameters of the model are determined by fitting
the pion-nucleus scattering data.
By using the pion-nucleus scattering wave functions generated from the
constructed optical potential and the electroweak pion production
amplitudes generated from the same SL model, we can calculate 
the coherent pion production cross sections.
 Our second step is to establish the reliability
of our model by comparing the predicted   
pion photo-production cross sections with the available data.
We then  proceed to calculate the neutrino-induced coherent pion production
cross sections.
In the following sections, we explain our calculational framework,
show some selected numerical results, and then conclude.

\section{A Dynamical Model}

We briefly explain how we calculate the amplitude
$A_{\lambda t\to \pi t}$ for the
coherent pion production off a nucleus in its ground state
($t$) induced by
an external current denoted by $\lambda$. 
Our starting point is a set of elementary amplitudes generated from the
SL model, $a^{SL}_{\lambda N\to \pi N} = a_\Delta + a_{nr}$, where 
$a_\Delta$ 
($a_{nr}$) denotes the resonant (non-resonant) part.
The resonant part can be written as
$a_\Delta = N/D(W)$, where
$D(W)=W-m_\Delta^0-\Sigma(W)$ with 
$W$, $m_\Delta^0$, $\Sigma$ denoting the $\pi N$ invariant mass, the bare
$\Delta$ mass, and the $\Delta$ self energy, respectively.
Within the $\Delta$-hole model\cite{karaoglu}, 
we can include the medium effects
by modifying the $\Delta$ propagator $D(W)$, and 
obtain the following expression of nuclear amplitude
$A_{\lambda t\to \pi t}$ 
\begin{eqnarray}
\eqn{mr2}
A_{\lambda t\to \pi t}
(\bm{k},\bm{q}) &=& \sum_{j} 
\int {d^3p_\Delta \over (2\pi)^3}
\psi_j^*(\bm{p}_N^\prime) 
\left[
{N(\bm{\tilde{k}},\bm{\tilde{q}}) \over
D(E+m_N-H_\Delta) - \Sigma_{\rm pauli} - \Sigma_{\rm spr} }
+ a_{nr}(\bm{\tilde{k}},\bm{\tilde{q}})
\right]
\psi_j(\bm{p}_N) \ ,
\end{eqnarray}
where $\bm{q}$ and $\bm{k}$ 
($\bm{\tilde{q}}$ and $\bm{\tilde{k}}$)
are the momenta for the external current and the
pion in the pion-nucleus (pion-nucleon) center-of-mass frame;
$\bm{p}_\Delta=\bm{p}_N+\bm{q}=\bm{p}_N^\prime+\bm{k}$.
A single nucleon wave function in the initial
[final] nucleus is denoted by
$\psi_j(\bm{p}_N)$ [$\psi_j(\bm{p}^\prime_N)$]
with the index $j$ specifying a nucleon orbit
including the isospin state.
The summation ($\sum_j$) is taken over the occupied states
of the nucleus.
The medium effects on the $\Delta$-propagator are
described by the $\Delta$ Hamiltonian ($H_\Delta$),
the Pauli-correction to the $\Delta$ self-energy ($\Sigma_{\rm pauli}$),
and the so-called spreading potential ($\Sigma_{\rm spr}$).
The spreading potential has the central and spin-orbit parts
with their strengths determined by two free complex parameters in our model.
The total energy of the pion-nucleus system is $E+Am_N$ 
($A$: mass number).
The integration over $\bm{p}_\Delta$ in Eq.(1)
can be done analytically 
by fixing the $\bm{p}_\Delta$-dependence of
the function in the square bracket to an effective constant value,
except for the kinetic term in $H_\Delta$ which is
is the source of the non-locality of the $\Delta$ propagation in
nuclei.
We then can sum over the nucleon states to obtain an expression which 
depends on the nuclear density;
the nuclear density has been determined experimentally.

A full transition amplitude is obtained by convoluting the amplitude in
\Eq{mr2} with the pion scattering waves. In this work,
the pion scattering wave functions are obtained
with an optical potential ($U_{\pi t}$).
Our starting point of constructing $U_{\pi t}$
is the $\pi N$ scattering amplitude generated from the SL model
$T_{\pi N}^{SL} = t_\Delta + t_{nr}$, where
 $t_\Delta$ and  $t_{nr}$ are the resonant and non-resonant amplitudes,
respectively. Consistently with \Eq{mr2}, the 
 $\Delta$-propagator of the resonant part 
$t_\Delta = F_{\pi N\Delta}F_{\pi N\Delta}/D(W)$ is modified by the
same procedure.
The resulting optical potential is given by
\begin{eqnarray}
\eqn{mr3}
U_{\pi t} (\bm{k}^\prime,\bm{k}) &=& \sum_{j} 
\int {d^3p_\Delta \over (2\pi)^3}
\psi_j^*(\bm{p}_N^\prime) 
\left[
{F_{\pi N\Delta}(\bm{\tilde{k}}^\prime)
F_{\pi N\Delta}(\bm{\tilde{k}})
\over
D(E+m_N-H_\Delta) - \Sigma_{\rm pauli} - \Sigma_{\rm spr} }
+ t_{nr}(\bm{\tilde{k}}^\prime,\bm{\tilde{k}})
\right]
\psi_j(\bm{p}_N) + c \rho^2 \ .
\end{eqnarray}
The last term of the above expression is a phenomenological 
term proportional to the
square of the nuclear density. It describes the absorption
of $s$- and $p$-wave pions by a pair of nucleons and
hence is determined by
two complex couplings $c_s$ and $c_p$.
Thus we have four free parameters in our model:  $c_s$ and $c_p$, and
two parameters specifying  the strength of the spreading
potential which are needed to calculate $\Sigma_{\rm spr}$ in Eqs.(1)-(2).
We determine these four parameters by fitting to 
the pion-nucleus scattering data. We then can  perform  parameter-free
calculations of the coherent electroweak pion production cross sections.

\section{Results}

\subsubsection{Pion-nucleus scattering}

\begin{figure}[t]
\begin{minipage}[t]{82mm}
 \begin{center}
 \includegraphics[width=82mm]{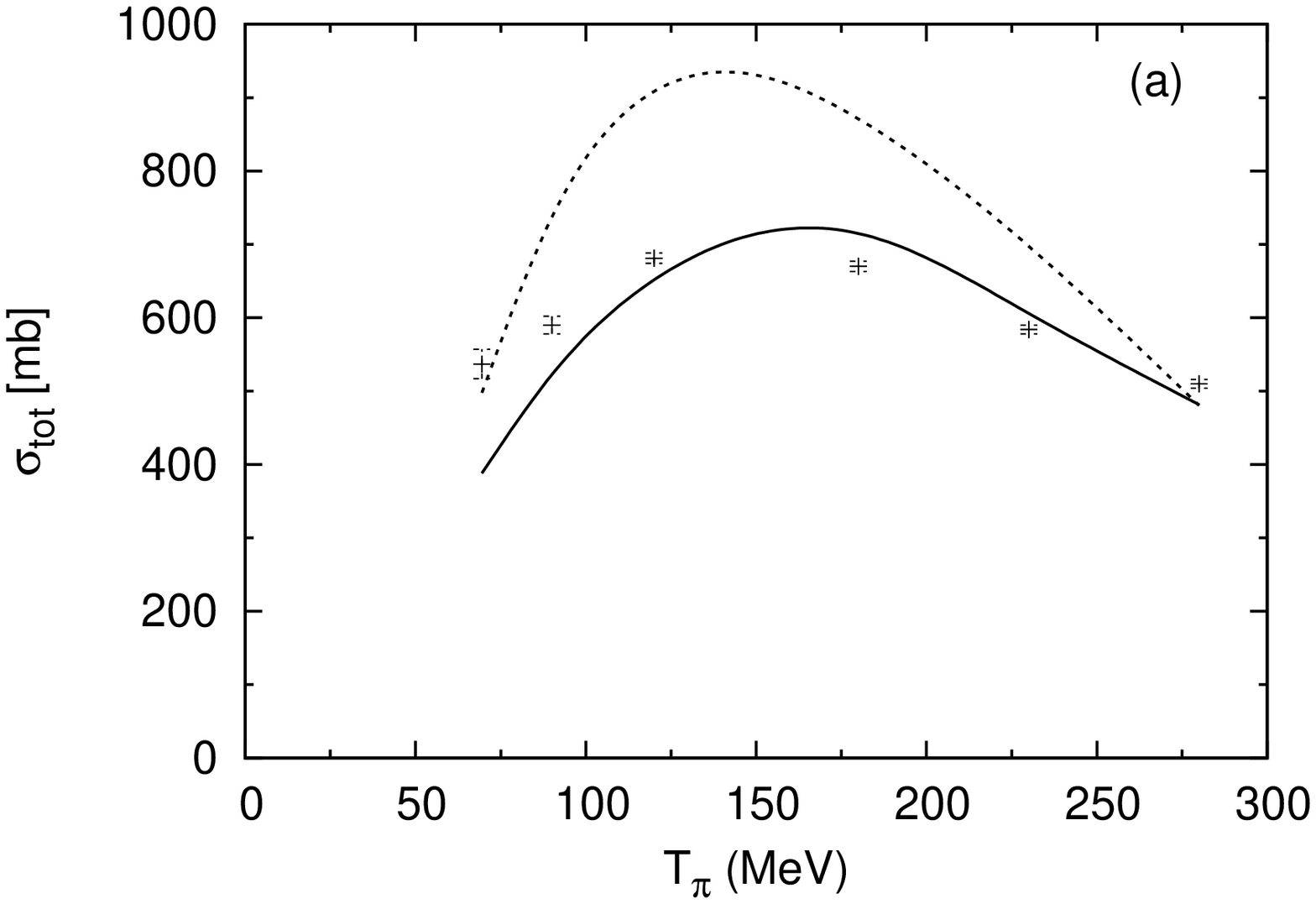}
\caption{\label{fig_piA} (a)
Total cross sections for $\pi^- - {}^{12}$C scattering. 
The solid curve is obtained with our full calculation, while the dashed
curve is obtained without the spreading potential.
(b) $\pi^- - {}^{12}$C elastic differential cross
 sections at $T_\pi$ = 180~MeV.
The solid curve is obtained with our full calculation.
In both figures, the data are from Ref.~\cite{pi-nucleus1}.
}
 \end{center}
\end{minipage}
\hspace{3mm}
\begin{minipage}[t]{82mm}
 \begin{center}
 \includegraphics[width=82mm]{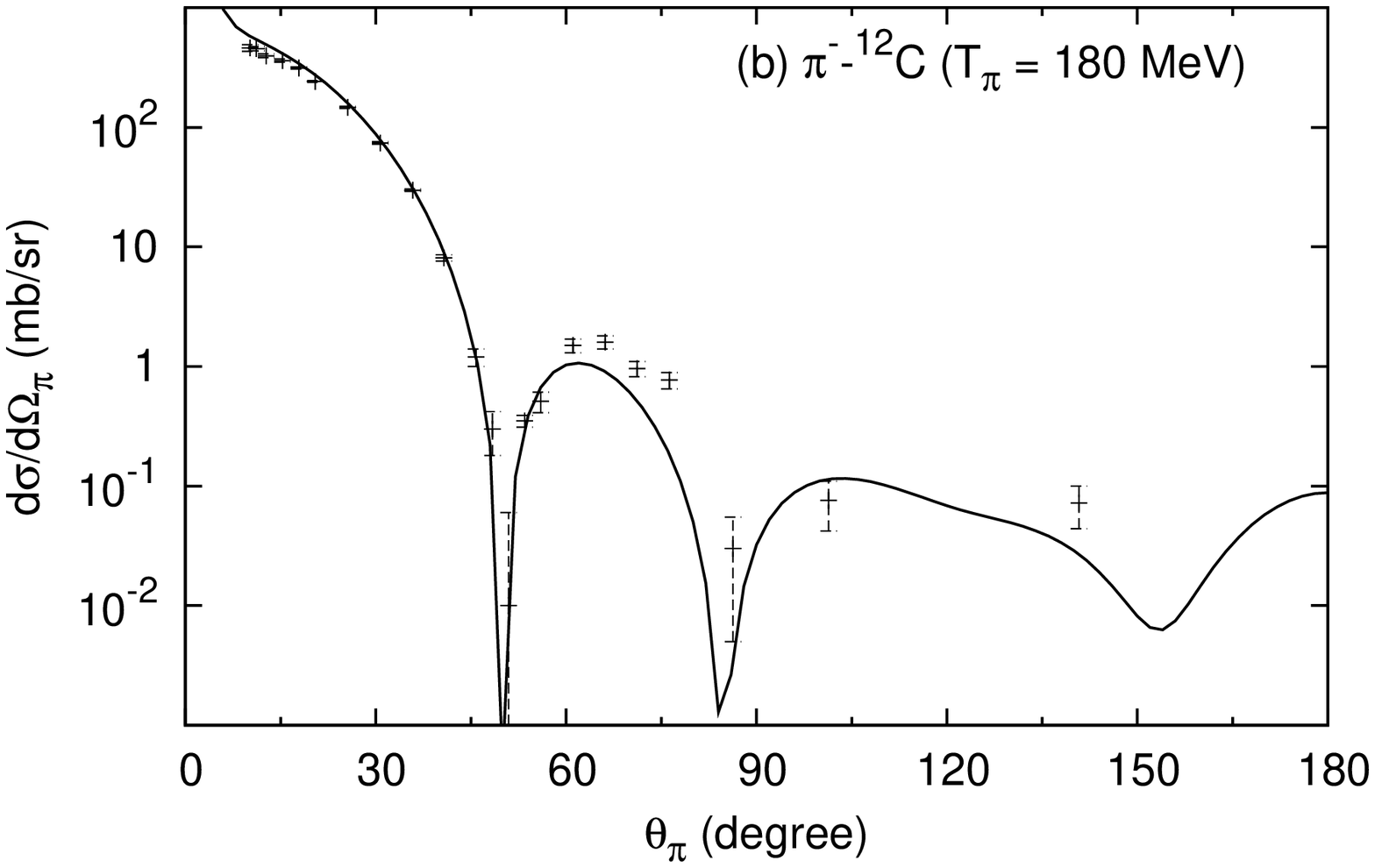}
 \end{center}
\end{minipage}
\end{figure}
We determine the four free parameters of our model
by fitting to the available  pion-nucleus scattering data
in the 50~MeV \ltap $T_\pi$ \ltap 300~MeV region.
($T_\pi$ is the pion kinetic energy in the laboratory frame.)
Two sample results from our fits to the $\pi-{}^{12}$C scattering data
are shown in Fig.~\ref{fig_piA}.
In the left-hand side of Fig.~\ref{fig_piA}, 
we see that the total cross sections for $\pi^-$-$^{12}$C
scattering as a function of $T_\pi$ can be reproduced very well
by our full calculations (solid curve).
For a comparison,
the results obtained without including  the spreading potential
are shown in the dashed curve.
We observe a large reduction 
 from the dashed to solid curves, indicating
the importance of the strong pion absorption simulated 
by the spreading potential.
In the right hand side of Fig.~\ref{fig_piA}, we see that
our model can also describe very well the differential cross sections.
Overall, our model can
satisfactorily reproduce the data 
for both the total and elastic cross sections in the considered
50~MeV \ltap $T_\pi$ \ltap 300~MeV energy region.

With the four parameters of our model determined in the fits to
the pion-nucleus scattering, we are
now in a position to perform parameter-free calculations of
the coherent electroweak pion production cross sections.

\subsubsection{Coherent pion photo-production}

\begin{figure}[t]
\begin{minipage}[t]{82mm}
 \begin{center}
 \includegraphics[width=82mm]{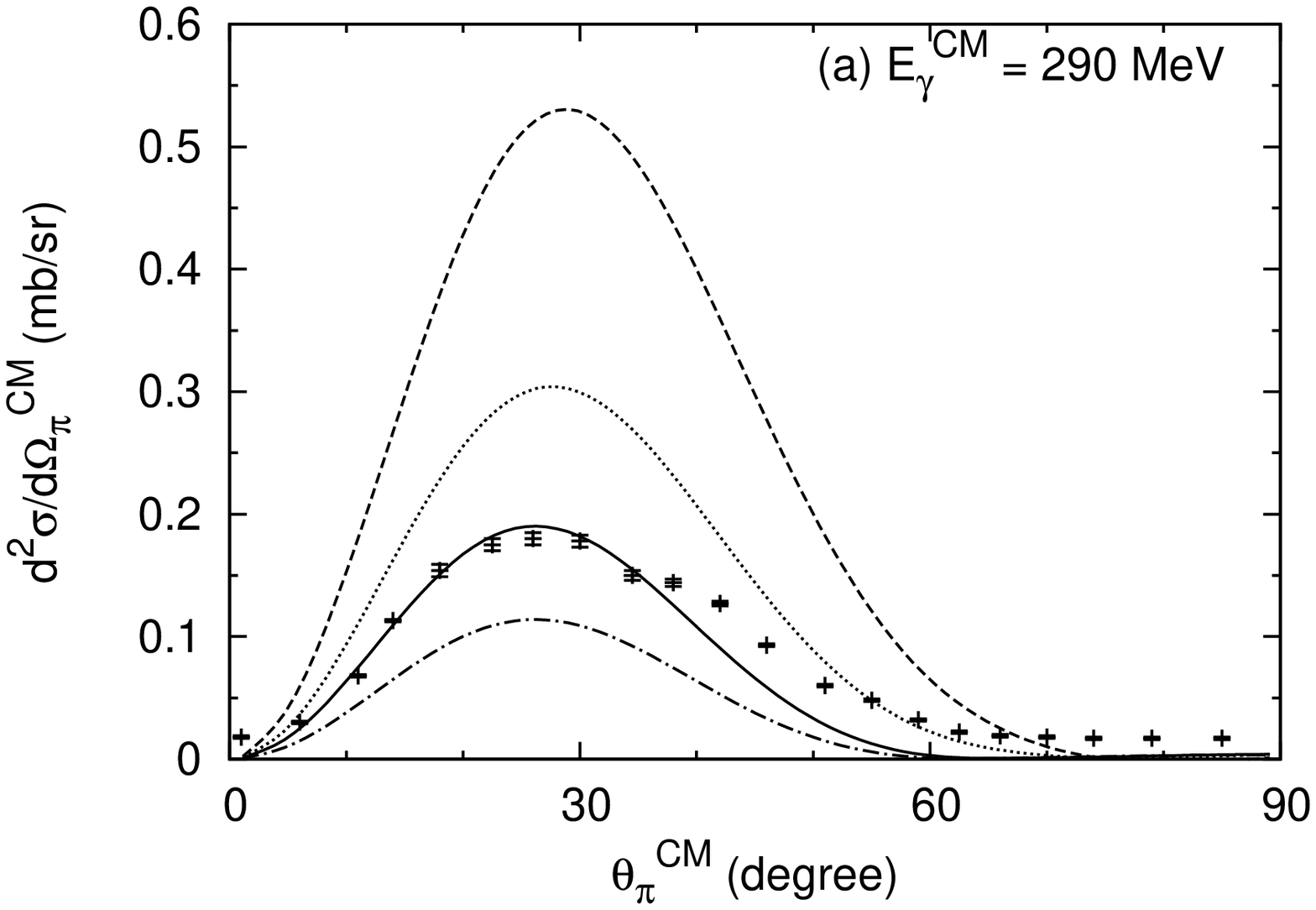}
\caption{\label{fig_krusche}
(a) Differential cross sections for 
$\gamma + {}^{12}{\rm C}_{g.s.} \to \pi^0 + {}^{12}{\rm C}_{g.s.}$.
The solid line represents the result of the full calculation.
The long-dashed line is obtained without the FSI and without
the medium effects on the $\Delta$-propagation,
while the short-dash line is obtained with the medium effects 
on the $\Delta$ included.
The dash-dotted line corresponds to
a case in which the pion production operator 
includes only the $\Delta$ mechanism.
The data are from Ref.~\cite{krusche}.
(b)
The $E_\nu$-dependence of the total cross section for
$\nu_\mu + {}^{12}{\rm C}_{g.s.} \to \mu^- + \pi^+ + {}^{12}{\rm
 C}_{g.s.}$ (solid line),
$\nu + {}^{12}{\rm C}_{g.s.} \to \nu + \pi^0 + {}^{12}{\rm C}_{g.s.}$
 (dashed line),
$\bar{\nu}_\mu + {}^{12}{\rm C}_{g.s.} \to \mu^+ + \pi^- + {}^{12}{\rm
C}_{g.s.}$ (dotted line) and
$\bar{\nu} + {}^{12}{\rm C}_{g.s.} \to \bar{{\nu}} + \pi^0 + {}^{12}{\rm C}_{g.s.}$ (dash-dotted line).
}
 \end{center}
\end{minipage}
\hspace{3mm}
\begin{minipage}[t]{82mm}
 \begin{center}
 \includegraphics[width=82mm]{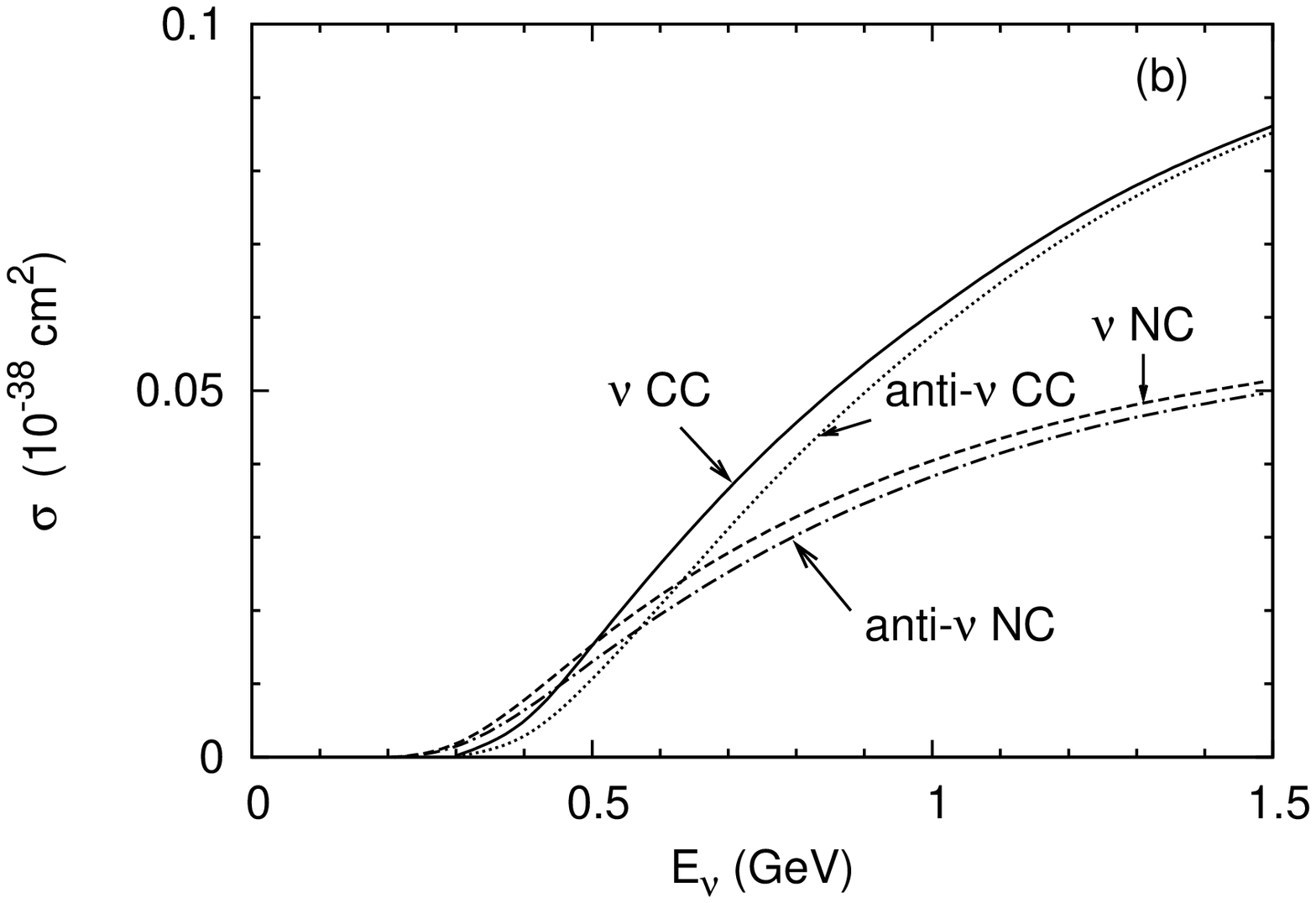}
 \end{center}
\end{minipage}
\end{figure}

The photo-production process, for which extensive data are available,
provides a good testing ground
for  our approach.
In the left-hand-side of Fig.~\ref{fig_krusche}
we compare the existing 
data with our results for the differential cross sections of
$\gamma + {}^{12}{\rm C}_{g.s.} \to \pi^0 + {}^{12}{\rm C}_{g.s.}$.
The solid curve is the result of our full calculation.
The long-dashed curve is obtained without including
the pion-nucleus final state interaction (FSI)  and without including the
the medium effects ($\Sigma_{\rm pauli}$ and $\Sigma_{\rm spr}$)
on the $\Delta$-propagation of Eq.(1).
When the the medium effects
on the $\Delta$ propagation in Eq.(1) is
included, we obtain the short-dash curve.
By comparing these three curves, 
it is clear that
the medium effects are quite sizable,
and they play an important role in bringing the calculated 
differential cross sections in agreement with the data.
The good general agreement seen in Fig.~\ref{fig_krusche}(a)
indicates the basic soundness of the method
we have used in determining the spreading potential.
It is true that, 
in the relatively larger angle region,
there are noticeable discrepancies
between the results of our full calculation and the data. 
We remark however that, as noted in Ref.~\cite{krusche},
the data in this region are likely to be substantially contaminated 
by incoherent processes in which the final
nucleus is in its low-lying excited states.

As a comparison, the dash-dotted curve in left hand side of
Figure~\ref{fig_krusche} is obtained
when the non-resonant pion production
amplitude\footnote{\label{footnote:non-res}
In the SL model, the resonant amplitude itself contains the non-resonant
mechanisms. We refer to purely non-resonant amplitudes as
``non-resonant amplitudes''.} is set to zero.
Clearly, the non-resonant production is substantial and can not
be neglected.
In general, we observe that, even near the resonance energy,
the contribution from
the non-resonant mechanism is quite significant.
This is partly because the resonant contribution is considerably
suppressed by pion absorption (the spreading potential) and the
non-local effect of $\Delta$ propagation ($\Delta$ kinetic term).

Through the comparison of our numerical results with the
pion photo-production data, we have established the reliability
of our approach based on a combined use of the SL model 
and the $\Delta$-hole model.
Thus we proceed to apply the same approach
to make predictions for cross sections 
of the neutrino-induced coherent pion production on $^{12}$C.

\subsubsection{Neutrino-induced coherent pion production}

\begin{figure}[t]
\begin{minipage}[t]{82mm}
 \begin{center}
 \includegraphics[width=82mm]{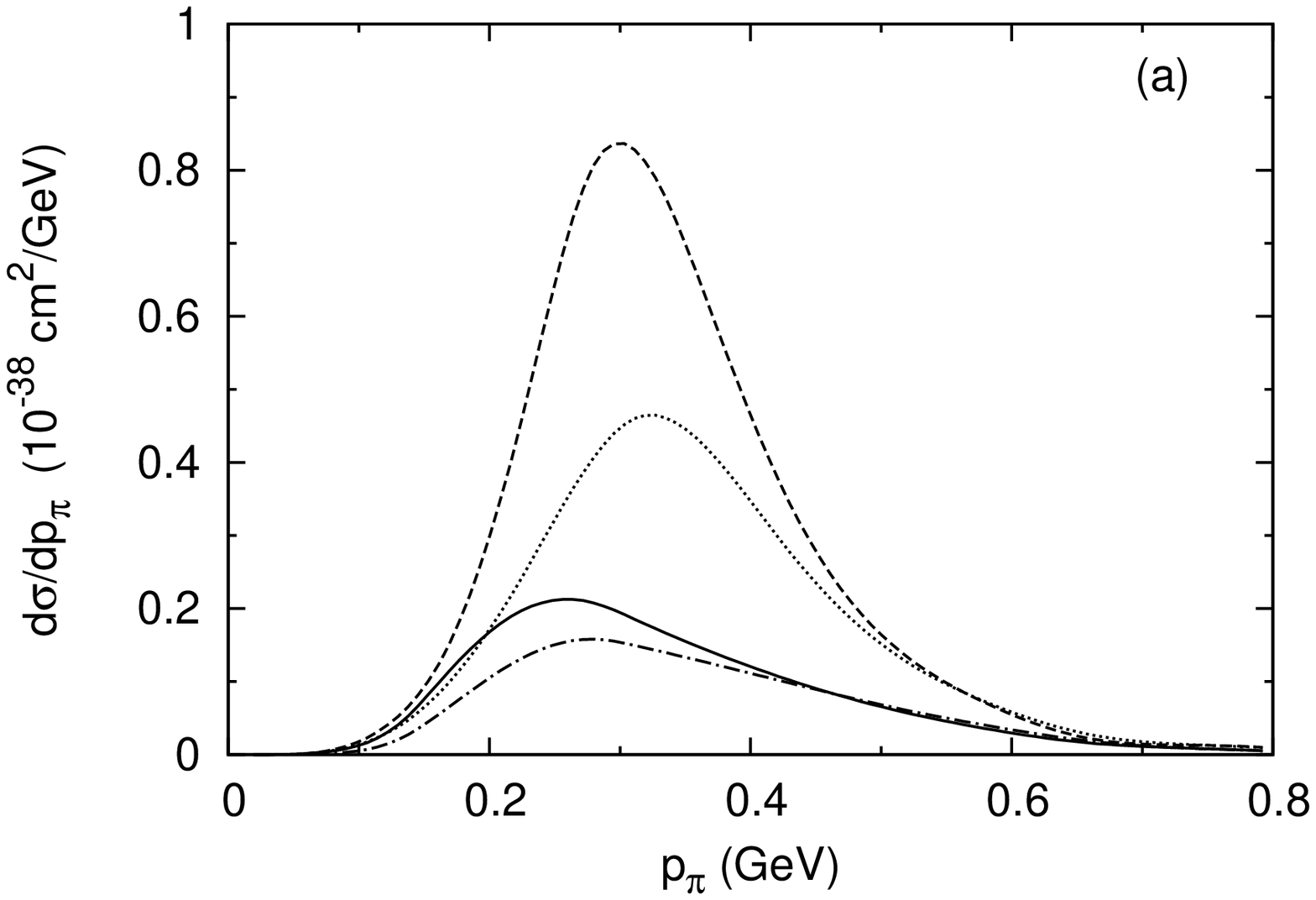}
\caption{\label{fig_pmom} (a)
The pion momentum distribution for
 $\nu_\mu +\! {}^{12}{\rm C}_{g.s.} \to \mu^- \!+ \!\pi^+\!+ \!{}^{12}{\rm
  C}_{g.s.}$ at $E_\nu$ = 1 GeV;
$p_\pi$ is the pion momentum in the laboratory frame.
The use of the solid, dashed, dotted and dash-dotted lines
follows the same convention as in Fig.~\ref{fig_krusche}(a).
(b)
The flux-convoluted $\eta$-distribution 
for $\nu + {}^{12}{\rm C}_{g.s.} \to \nu + \pi^0 + {}^{12}{\rm C}_{g.s.}$ obtained in our full calculation.
The neutrino flux is taken from MiniBooNE~\cite{miniboone_flux}.
Also shown is the Monte Carlo result 
from MiniBooNE~\cite{miniboone} rescaled.
}
 \end{center}
\end{minipage}
\hspace{3mm}
\begin{minipage}[t]{82mm}
 \begin{center}
 \includegraphics[width=82mm]{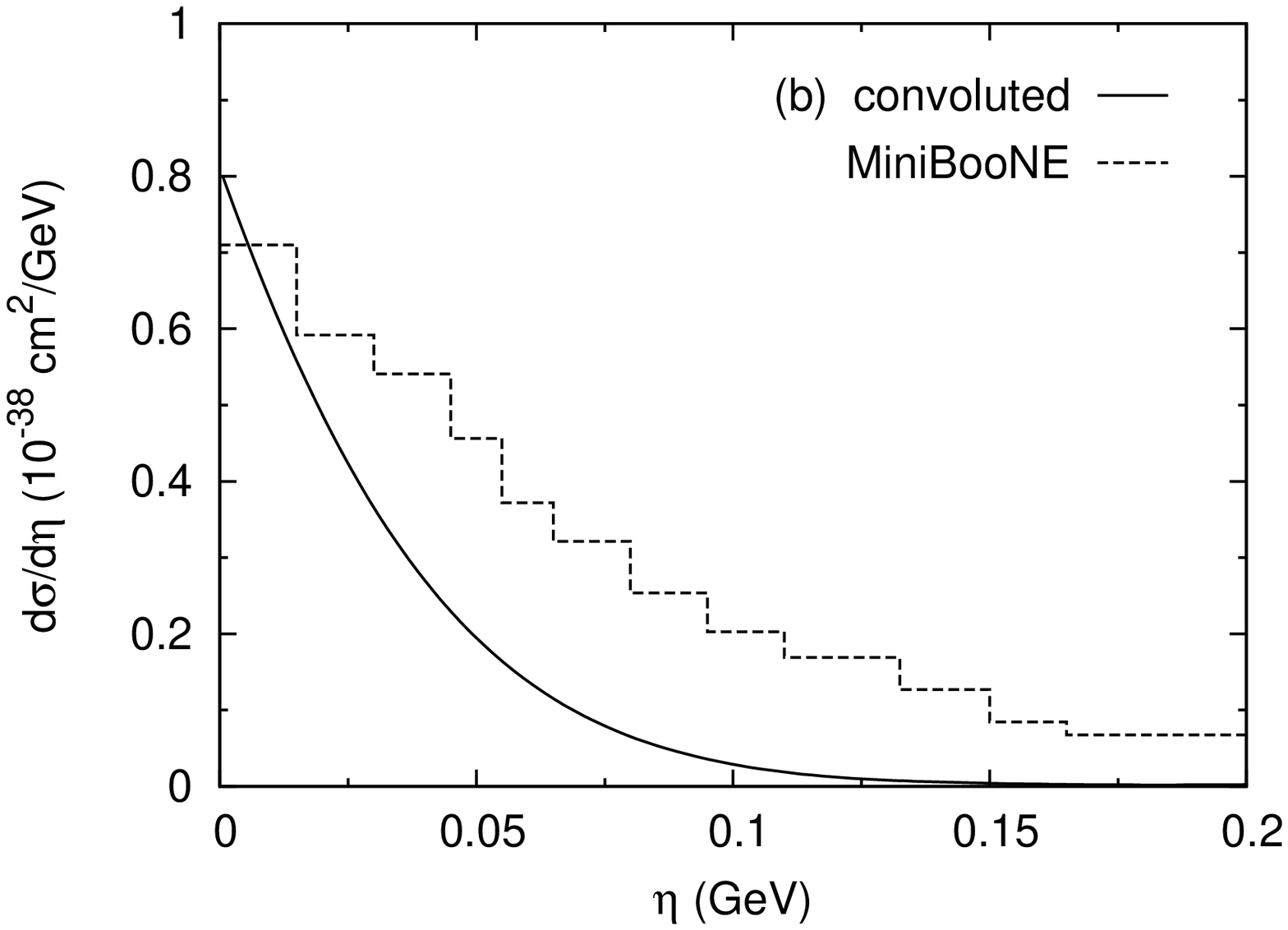}
 \end{center}
\end{minipage}
\end{figure}

We now present our predictions of
the neutrino-induced coherent pion production
on the $^{12}$C target.
We consider the CC and NC processes induced by
a neutrino or an anti-neutrino.
In the right-hand-side of Figure~\ref{fig_krusche}.
we show  the predicted total cross sections for these processes
as functions of the incident neutrino (anti-neutrino) energy $E_\nu$
in the laboratory system.
It is seen that, for higher incident energies, 
the ratio $\sigma_{CC}/ \sigma_{NC}$ approaches 2,
a value expected from the isospin factor.
For lower incident energies ($E_\nu$\ltap 500~MeV), however, 
$\sigma_{NC}$ is larger than $\sigma_{CC}$,
reflecting the fact that the phase space for the CC process
is reduced significantly by the muon mass.
To compare our results with data,
we need to evaluate the total cross sections averaged
over the neutrino fluxes that pertain to the relevant experiments.
We choose to use the fluxes up to $E_\nu \le$ 2 GeV 
and neglect the fluxes beyond that limit.
We also need to consider the experimental setting such as kinematical cuts.
A K2K experiment~\cite{hasegawa} reports the upper limit for the
neutrino CC coherent pion production,
$\sigma_{\rm K2K}  <  7.7 \times 10^{-40} {\rm cm}^2$.
This upper limit corresponds to events satisfying
the muon momentum cut, $p_\mu > 450$ MeV and the cut on the
square of the momentum
transfer, $Q_{\rm rec}^2 < 0.1$ GeV$^2$; 
$Q_{\rm rec}$ is the momentum transfer
reconstructed with an
assumption of the quasi-free kinematics.
Therefore, we also 
calculate the total cross section with these cuts and then 
convolute with the flux
reported by the K2K experiment~\cite{K2K}.
We obtain $\sigma_{\rm ave}^{CC} = 6.3 \times 10^{-40} {\rm cm}^2$ 
which is consistent with the upper limit from K2K.

For the NC process, we use the flux reported 
by MiniBooNE in Ref.~\cite{miniboone_flux} and arrive at
$\sigma_{\rm ave}^{NC} = 2.8 \times 10^{-40} {\rm cm}^2$;
no kinematical cut is necessary in this case.
This is to be compared with
$\sigma_{\rm MiniBooNE}  =  7.7 \pm 1.6\pm 3.6 \times 10^{-40} {\rm cm}^2$
given in Ref.~\cite{raaf}.
Our result is consistent with the empirical value 
within the large experimental errors, even though 
the theoretical value is rather visibly smaller 
than the empirical central value.
It is to be noted however that Ref.~\cite{raaf}
is a preliminary report,  
and that, as discussed in great detail in Ref.~\cite{amaro}, 
$\sigma_{\rm MiniBooNE}$ may be overestimated 
due to the use of the Rein-Sehgal (RS) model\cite{RS} in the analysis.

Next, we present our result for
the pion momentum spectrum for the CC process
at $E_\nu$ = 1 GeV
in Fig.~\ref{fig_pmom}(a).
The importance of the medium effects 
manifests itself here in the same manner as 
in the photo-process [Fig.~\ref{fig_krusche}(a)].
In the $\Delta$ region, strong pion absorption 
is seen to reduce the cross sections significantly,
and FSI shifts the peak position.
The dash-dotted line corresponds to a case
in which the pion production operator
contains only the $\Delta$ mechanism 
(without non-resonant contributions), 
while the pion optical potential is kept unchanged. 
We note that, at $E_\nu$ = 1 GeV,
the dash-dotted line corresponds to 
82\% of the solid line,
and for $E_\nu$ = 0.5 GeV, 64\%.
The result indicates that the non-resonant components 
in our model play a significant role in the coherent pion production; 
their role is particularly important for $E_\nu$\ltap 0.5 GeV.
This characteristic feature of our model 
should be contrasted with the fact that non-resonant mechanisms
play essentially no role in any of the previous microscopic calculations
for neutrino-induced coherent pion production.

We examine the effects of the non-locality of 
$\Delta$-propagation in nuclei arising from 
the $\Delta$ kinetic term.
For this purpose, we introduce a quantity
${\cal R}(E_\nu)\equiv \sigma(E_\nu)/\sigma_{\rm local}(E_\nu)$,
where $\sigma$ ($\sigma_{\rm local}$) is calculated with (without)
the $\Delta$ kinetic term.
This subject has been studied in Ref.~\cite{non-local}
which included the $\Delta$-mechanism only, without FSI or medium
modifications of the $\Delta$.
They showed that the cross section is changed by
${\cal R}$ \ltap 0.5, $\sim$ 0.6, \ltap 1 at $E_\nu$ = 0.5, 1, 1.5 GeV.
When we use the free $\Delta$-propagator without FSI,
which contains the physics similar to that of Ref.~\cite{non-local},
we find that the non-locality change the cross sections by
${\cal R} =$ 0.41, 0.74, 0.88  at $E_\nu$ = 0.5, 1, 1.5 GeV.
This result is fairly close to that obtained in Ref.~\cite{non-local}.
Next we use the $\Delta$-propagator in \Eq{mr2}
but without the major medium effects
($\Sigma_{\rm spr}$, $\Sigma_{\rm pauli}$, and small effects),
and then find that the effect of the non-locality is somewhat moderated:
${\cal R} =$ 0.54, $1.03$ and $1.14$
at $E_\nu$ = 0.5, 1.0 and 1.5 GeV.
%
%
Including the non-resonant amplitudes, medium effects on the $\Delta$
and FSI,
the effect of the non-locality is
${\cal R} =$ 0.34, 0.73, 0.84  at $E_\nu$ = 0.5, 1, 1.5 GeV
for the full
calculation, showing that
the non-locality is still important.
Even though the non-locality might be partly accounted for
with the use of the spreading potential fitted to data, considering its
importance,
it is preferable to take it into account explicitly.

%
The MiniBooNE analysis of NC data used the $\eta$-distribution
[\,$\eta \equiv E_\pi (1-\cos\theta_\pi)$]
to distinguish coherent pion production from the other processes
contributing to the $\pi^0$-production events.
To this end, MiniBooNE used the ``shape`` of the $\eta$-distribution 
obtained from the RS model~\cite{RS}
with the momentum reweighting function applied.
However, a microscopic calculation in Ref.~\cite{amaro}
was found to give an $\eta$-distribution
appreciably different from that obtained in the RS model,
and the authors of Ref.~\cite{amaro}
pointed out a possibility that the MiniBooNE
might have substantially overestimated the NC events.
Thus it is interesting to compare our result with the
$\eta$-distribution from MiniBooNE.
Figure~\ref{fig_pmom}(b) shows the ``average'' $\eta$-distribution,
resulting from the convoluting of the $\eta$-distribution 
obtained in our present calculation 
with the MiniBooNE neutrino flux~\cite{miniboone_flux}. 
The figure also presents
the MiniBooNE Monte Carlo results 
({\it cf.} Fig.~3b of Ref.~\cite{miniboone}),
arbitrarily rescaled to match the theoretical curve
at $\eta$ = 0.005 GeV.
We remark that
the $\eta$-distribution we have obtained 
is fairly close to that given in Ref.~\cite{amaro}.
and thus also arrive at the conclusion
that it is possible that MiniBooNE
might have substantially overestimated the NC events.

\section{Discussion and summary}

Recent theoretical efforts on the neutrino-induced coherent pion
production are indeed active, and 
it is interesting to see
a numerical comparison of those calculations.
Such a comparison has been presented at this
conference\cite{comparison_nuint09}.
Because of the lack of data, it is impossible to rate the models from the
comparison.
In the following, however, we argue that our present calculation would
have advantages over the existing calculations.
Our model provides, thanks to the consistency built in
the SL model, a consistent description of the
pion-nucleus scattering and the photo- or neutrino-induced coherent pion
productions. 
Because of the consistency, we were able to fix the parameters using the
 data for the pion-nucleus scattering and to
 predict the coherent processes.
It is also the consistency which enables us to
 test our approach using the
existing data for the photo-process.
No other microscopic calculation for the neutrino-induced coherent pion
production has been checked to this level.
Besides, all the other previous
calculations lack the non-locality of
$\Delta$-propagation whose importance was examined in the previous
section.
Further remark is concerned with FSI. Our optical potential is made to
reproduce the data for the pion-nucleus scattering in the range of 
50~MeV \ltap $T_\pi$ \ltap 300~MeV, covering the most important
$\Delta$ region.
On the other hand, in the other microscopic calculations, their optical
potential may not be as accurate as ours. 
For example, let us look into the model of Amaro et al.~\cite{amaro},
which is the most sophisticated among the existing microscopic models 
for neutrino-induced coherent pion production.
The parameters in their optical potential
were fitted 
to the data for the binding energies 
of pionic atoms.
Because, in the present context, the optical potential should
work well over relatively wide energy region around the
$\Delta$ region, 
the optical potential used in 
Ref.~\cite{amaro} may contain a questionable aspect.
Finally we mention that it has become fairly clear that
the RS model, the prominent PCAC-based model, does not work reasonably
 for $E_\nu$\ltap 2 GeV.
This point was made clear through the detailed study of
the relation between the RS model and the microscopic model 
in Ref.~\cite{amaro}. 
Although some improvements on the RS model has been
proposed\cite{bs_pcac,paschos,hernandez},
the applicability of a PCAC-based model to the relevant energy region
is currently a subject of controversy.

In summary, we have developed a microscopic model based on a combination
of the SL and the $\Delta$-hole models.
The model provides a consistent description of the pion-nucleus
scattering and the coherent pion production processes.
Utilizing the consistency, we can fix the free parameters in our model
using the pion-nucleus scattering data, and then give
parameter-free predictions for the coherent processes. 
After testing the reliability of our approach using the data for the
photo-process, we calculate and present results for the neutrino-induced
coherent pion production.


\begin{theacknowledgments}
This work is supported by
the Natural Sciences and Engineering Research Council of Canada
and Universidade de S\~ao Paulo (SXN),
by the U.S. Department of Energy, Office of Nuclear Physics,
under contract DE-AC02-06CH11357 (TSHL),
by the Japan Society for the Promotion of Science,
Grant-in-Aid for Scientific Research(C) 20540270 (TS),
and by the U.S. National Science Foundation
under contract PHY-0758114 (KK).
\end{theacknowledgments}

\end{document}